\documentclass{article}
\usepackage{spconf,amsmath,graphicx}
\usepackage{amsfonts}
\usepackage{booktabs}
\usepackage{algorithm}
\usepackage{algpseudocode}
\usepackage{listings}

\title{Music Source Separation with Band-Split RoPE Transformer}
\name{Wei-Tsung Lu$^*$, Ju-Chiang Wang$^*$, Qiuqiang Kong, Yun-Ning Hung \thanks{$^*$ Equal contribution}}
\address{SAMI, ByteDance\\
\tt\small \{weitsung.lu, ju-chiang.wang, kongqiuqiang, yunning.hung\}@bytedance.com
}
\begin{document}
\ninept
\maketitle
\begin{abstract}
Music source separation (MSS) aims to separate a music recording into multiple musically distinct stems, such as vocals, bass, drums, and more. 
Recently, deep learning approaches such as convolutional neural networks (CNNs) and recurrent neural networks (RNNs) have been used, but the improvement is still limited. 
In this paper, we propose a novel frequency-domain approach based on a Band-Split RoPE Transformer (called BS-RoFormer). BS-RoFormer relies on a band-split module to project the input complex spectrogram into subband-level representations, and then arranges a stack of hierarchical Transformers to model the inner-band as well as inter-band sequences for multi-band mask estimation. To facilitate training the model for MSS, we propose to use the Rotary Position Embedding (RoPE). The BS-RoFormer system trained on MUSDB18HQ and 500 extra songs ranked the first place in the MSS track of Sound Demixing Challenge (SDX'23). Benchmarking a smaller version of BS-RoFormer on MUSDB18HQ, we achieve state-of-the-art result without extra training data, with 9.80 dB of average SDR.

\end{abstract}
\section{Introduction}
\label{sec:intro}

Music source separation (MSS) \cite{rafii2018overview, mitsufuji2022music} is a task of separating a music recording into musically distinct sources. As defined in the 2015 Signal Separation Evaluation Campaign (SiSEC) \cite{liutkus20172016}, researchers are focused on the 4-stem setting: vocals, bass, drums, and other. 
The MUSDB18 dataset \cite{rafii2017musdb18} has been used to benchmark the performance. 
MSS is considered to be a more challenging audio separation task given  the complexity of input signals (i.e., stereo, 44.1k Hz sampling rate) and more sources to separate (i.e., 4 or more stems). MSS can benefit various downstream MIR tasks, such as vocal pitch estimation \cite{nakano2019joint}, music transcription \cite{lin2021unified}, and so on. MSS may also enable applications such as karaoke-version generation, intelligent music editing and remixing \cite{ripple}, and other music inspired derivative works.

In recent years, many deep learning approaches have been proposed to tackle the MSS problem. They are typically categorized into frequency-domain and time-domain approaches. Frequency-domain approaches reply on Fourier transform to derive a time-frequency (T-F) representation for input. Then, models such as fully connected neural networks \cite{grais2014deep}, convolutional neural networks (CNNs) \cite{chandna2017monoaural, kong2021decoupling, jansson2017singing}, and recurrent neural networks (RNNs) \cite{uhlich2017improving} are applied. 
On the other hand, time-domain approaches such as Wave-U-Net \cite{stoller2018wave}, ConvTasNet \cite{luo2019conv}, and Demucs \cite{defossez2019music} build their neural networks directly on the waveform input. Recently, Hybrid Transformer Demucs (HTDemucs) \cite{rouard2023hybrid} proposes to use a cross-domain Transformer to combine the frequency- and time-domain models.

Speaking of frequency-domain approaches, most existing models do not make assumptions on weighting the frequency bins based on prior knowledge. Models like CNNs use the same kernels across all frequency bins for all target sources, expecting the models can learn the band-pass mechanism from the data with raw frequency bins.
However, this might not be efficient, since different frequency bands may have different patterns that preferably characterize different instrument sources.
Recently, band-split mechanism is introduced at the front-end in Band-split RNN (BSRNN) \cite{luo2023music} to force the model to learn band-wise features. BSRNN splits the T-F representation into non-overlapping subbands and arranges a stack of interleaved RNNs to simultaneously model the inner-band and inter-band sequences. This design has demonstrated state-of-the-art result on Musdb18. However, its use of RNN could be still sub-optimal, given that Transformer \cite{vaswani2017attention} has constantly shown the superiority in many relevant tasks of sequential data modeling.


The idea of modeling the T-F representation of music audio using interleaved Transformers has been explored recently \cite{lu2021spectnt, lu2023multitrack}. For example, SpecTNT \cite{lu2021spectnt} consists of two Transformer encoders in a hierarchical order to model the frequency and time sequences respectively, and then uses the frequency class tokens to connect between the two Transformers. This or similar ideas have shown state-of-the-art performance in various music transcription related tasks \cite{lu2023multitrack, hung2022modeling, wang2022catch}, but have not been studied for MSS.
In this paper, we propose a novel MSS approach based on Band-Split RoPE Transformer (termed as \emph{BS-RoFormer}) that originates from combining the ideas of band-split and hierarchical Transformer architecture. Making it work, however, is non-trivial, mainly because the model is large, and training it effectively is very memory and time consuming. We introduce the use of Rotary Position Embedding (RoPE) \cite{su2021roformer} in Transformer significantly improves the performance. Other efforts such as checkpointing, mixed precision, and flash attention \cite{dao2022flashattention} could also help the training efficiency. 
We submitted the BS-RoFormer system to Sound Demixing Challenge 2023 (SDX'23)\footnote{\scriptsize{https://www.aicrowd.com/challenges/sound-demixing-challenge-2023/}}, the Music Separation track. Our system ranked the first place and outperformed the second best by a large margin in SDR. 
In ablation study, we demonstrate the importance of RoPE and that a smaller BS-RoFormer model trained solely on MUSDB18HQ can achieve state-of-the-art performance compared to existing models.

\begin{figure*}[t]
  \centering
  \centerline{\includegraphics[width=\textwidth]{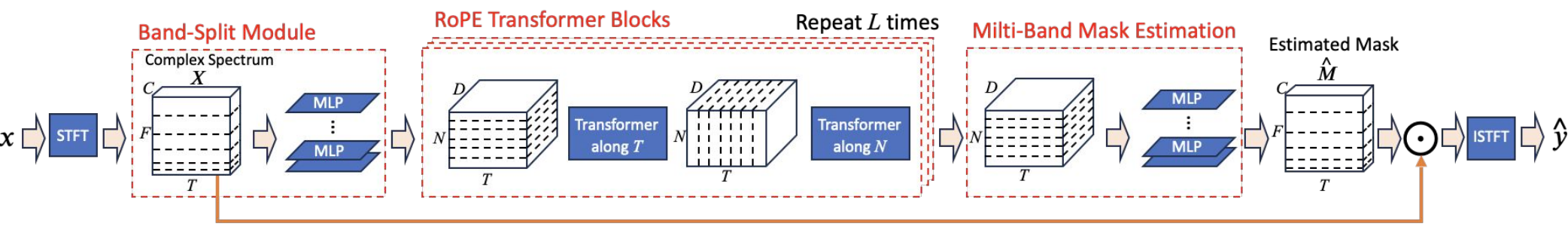}}
  \caption{The framework of the Band-Split RoFormer system.}
  \label{fig:bstransformer_framework}
\end{figure*}



\section{System Overview}
\label{section:freq_mss}

Fig. \ref{fig:bstransformer_framework} depicts the system.
Let $ x \in \mathbb{R}^{C \times L} $ denote the audio waveform of the input mixture, where $ C $ and $ L $ are the numbers of channels and audio samples, respectively. 
Our frequency-domain MSS system uses the complex spectrogram as input, where $ x $ is transformed into a time-frequency (T-F) representation $ X \in \mathbb{C}^{C \times T \times F} $ by a short-time Fourier transform (STFT), where $ T $ and $ F $ are the the numbers of frames and frequency bins, respectively. 
Let $ f_{\theta} $ denote a T-F neural network with a set of learnable parameters $ \theta $, the output of $ f_{\theta} $ can be linear spectrums, ideal binary masks (IBMs), ideal ratio mask (IRMs), or complex IRMs (cIRMs) \cite{wang2018supervised}. 
We adopt the cIRMs, denoted as $ \hat{M} \in \mathbb{C}^{C \times T \times F} $. In this work, the goal of $ f_{\theta} $ is to estimate the cIRMs: $ \hat{M} = f_{\theta}(X) $, so that both the magnitude and phase signals of the target source are attained. 

Then, the separated complex spectrogram $ \hat{Y} \in \mathbb{C}^{C \times T \times F} $ is obtained by multiplying the cIRM by the input complex spectrogram: $ \hat{Y} = \hat{M} \odot X $. Finally, an inverse STFT (iSTFT) is applied to $ \hat{Y} $ to recover the separated signal $ \hat{y} $ in the time-domain. Mean absolute error (MAE) loss between the reference $ y $ and output $ \hat{y} $ is used to train $ f_{\theta} $. Specifically, the objective loss includes both the time-domain MAE and the multi-resolution complex spectrogram MAE \cite{guso2022loss}:
\begin{equation} \label{eq:loss}
\text{loss} = || y - \hat{y} || + \sum_{s=0}^{S - 1}|| Y^{(s)} - \hat{Y}^{(s)}||,
\end{equation}
where $S=5$ multi-resolution STFTs are used with the window sizes of $[4096, 2048, 1024, 512, 256]$ and a fixed hop size of $ 147 $, which is equivalent to 300 frames per second.


\section{Band-Split RoPE Transformer}
\label{section:bs_transformer}

BS-RoFormer consists of a band-split module, RoPE Transformer blocks, and a multi-band mask estimation module.

\subsection{Band-Split Module}
In BSRNN \cite{luo2023music}, it has shown that the band-split strategy can help the frequency-domain approach. One rationale is that the input mixture audio $ X $ is a full-band signal,
so splitting the bands based on a prior setting can foster the model to purify the learned representations at different bands, gaining the robustness against cross-band vagueness.
Following \cite{luo2023music}, we split $ X $ into $ N $ uneven non-overlapping subbands along the frequency axis and apply individual multi-layer perceptions (MLPs) to each subband. We denote the output of each subband as $ X_{n} \in \mathbb{C}^{C \times T \times F_{n}} $, where $ F_{n} $ is the number of frequency bins in the $n$-th subband. All subbands $ X_{n} $ constitute the entire complex spectrum $ X $, and there is $ \sum_{n=1}^{N} F_{n} = F $. 

Each MLP consists of a RMSNorm layer \cite{zhang2019root} followed by a linear layer. 
The RMSNorm layer regularizes the summed inputs to a neuron in one layer according to root mean square. RMSNorm is an efficient replacement to the LayerNorm normalization \cite{zhang2019root}. The $n$-th linear layer contains a learnable matrix with a shape of $ (C \times F_{n}) \times D $ and a learnable bias with a shape of $ D $, where $ D $ is the number of features. The transformed output of each subband is denoted as $ H_{n}^{0} $ with a shape of $ T \times D $. We stack all $ H_{n}^{0}, n=1, \dots, N $, along the subband axis to obtain a stacked $ H^{0} $ with a shape of $ T \times N \times D $ as the input to the RoPE Transformer blocks.





\subsection{RoPE Transformer Blocks}

\begin{figure}[t]
  \centering
 \centerline{\includegraphics[width=0.8\columnwidth]{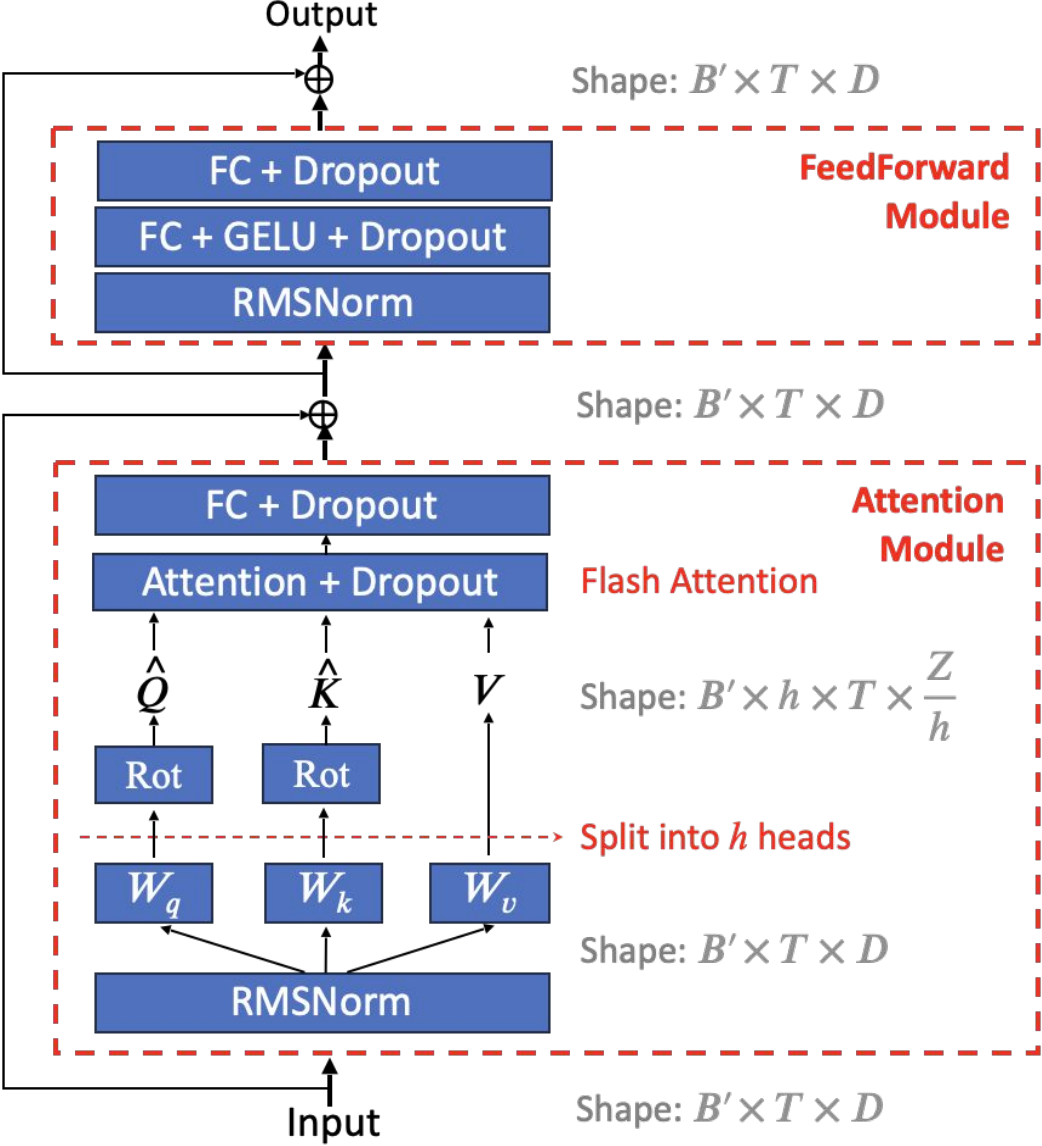}}
  \caption{Diagram of a Transformer layer. This example is a time-Transformer along the temporal sequence and $ B'=B \times N $.}
  \label{fig:transformer_layer}
\end{figure}

Suppose there are $ L $ blocks in RoPE Transformer blocks, each block's output is denoted as $ H_{n}^{l} \in \mathbb{R}^{T \times N \times D} $. The band-split output $ H_{n}^{0} $ serves as the input for the 1-st Transformer block.
Different from conventional Transformer encoder that applies self-attention to one-dimensional sequence (e.g., time), our Transformer structure is hierarchical, where interleaved Transformer layers are in turn applied to the time and frequency axes in a Transformer block. The former Transformer, called \emph{time-Transformer}, models the inner-band (local) temporal sequence, while the latter Transformer, called \emph{subband-Transformer}, handles the inter-band (global) spectral sequence to ensure the information across bands is exchangeable. 
The process of a Transformer block is presented in Algorithm 1. For time-Transformer, all mini-batches (with a batch size of $ B $) and subbands are stacked; for subband-Transformer, all mini-batches and frames are stacked. All Transformer layers use the same architecture. Fig. \ref{fig:transformer_layer} depicts a Transformer layer, which consists of an attention module (with rotary position embedding) and a feedforward module.




\begin{algorithm}[t]
  \caption{Hierarchical Transformer for a time-frequency input, with parameters $B$: batch size, $T$: frames number, $N$: subband number, and $D$: latent dimension.}\label{alg:transformer_block}
  \begin{algorithmic}[1]
        \item \textbf{Input}: $ H^{l} $ with a shape of $B \times T \times N \times D$
        \item Permute $ H^{l} $ to $(B \times N) \times T \times D$
        \item Apply a Transformer along $ T $.
        \item Rearrange the above output to $(B \times T) \times N \times D$
        \item Apply a Transformer along $ N $.
        \item \textbf{Output}: Permute the above output to $H^{l+1}$ with a shape of $B \times T \times N \times D$.
  \end{algorithmic}
  \label{alg:interleaved_transformers}
\end{algorithm}

\subsubsection{Attention Module with Rotary Position Embedding}
\label{sec:attention}

In an attention module, we first rearrange the input to a shape as needed (see Algorithm 1), 
followed by a RMSNorm \cite{zhang2019root}. Then, we apply the query, key, and value layers to predict the query, key, and value \cite{vaswani2017attention}: $Q=H^{l}W_{q}$, $K=H^{l}W_{k}$, and $V=H^{l}W_{v}$,
where $ Q $, $ K $, and $ V $ have the shapes of $ D \times Z $, and $ Z $ is the latent dimension. Learnable matrices $ W_{q} $, $ W_{k} $, and $ W_{v} $ have the shapes of $ D \times Z $. The $ Q $, $ K $, and $ V $ are split into multiple heads. 
For positional encoding, we propose to use the rotary position embedding (RoPE) \cite{su2021roformer},
which are applied as:
\begin{equation} \label{eq:rope}
\begin{split}
& \hat Q=\text{Rot}(Q) \\
& \hat K=\text{Rot}(K) 
\end{split}
\end{equation}
where $ \text{Rot}(\cdot) $ is the RoPE encoder \cite{su2021roformer} shared throughout the entire Transformer blocks. The time-Transformer and subband-Transformer have their own RoPE encoders that apply rotation matrices on each embedding in $Q$ and $K$ based on its position of the corresponding sequence.
Then, we apply the attention operation by using the processed query, key, and value as follows \cite{vaswani2017attention}:
\begin{equation} \label{eq:attention}
\text{output} = Dropout\left(Softmax(\frac{\hat{Q}\hat{K}^T}{\sqrt{Z/h}})\right)V,
\end{equation}
where $h$ is the number of heads. Equation (\ref{eq:attention}) is the core part that requires the most computation. To speed up, we employ the FlashAttention technique \cite{dao2022flashattention}. After the attention module, a fully-connected layer with Dropout is used. A residual addition is applied between the input and output of the attention module.




We believe the use of RoPE is crucial in this work. Given the proposed hierarchical Transformer structure, the positional embeddings for the time- and subband-Transformers should be robust to the alternating self-attention operations between the time and subband axes. Directly adding learnable absolute positional embeddings to encode the positions may fail to maintain the scale of norm after repetitive transposed self-attention processes. Since RoPE encodes the relative position by multiplying the context representations with a rotation matrix, we argue this would help preserve the positional information for both time and subband sequences.


\subsubsection{Feedforward Module}
The feedforward module consists of a RMSNorm layer, a fully connected layer with GeLU activation \cite{hendrycks2016gaussian}, and a dropout. Then, one more fully connected layer with dropout is applied. Similarly, we apply a residual addition between the input and output of the feedforward module. 


\subsection{Multi-band Mask Estimation Module}
We denote the outputs from the RoPE Transformer blocks as $ H_{n}^{L}, n=1, \dots, N$. Similar to the band-split module, we apply $ N $ individual MLP layers to each subband $ H_{n}^{L} $. Each MLP layer consists of a RMSNorm layer, a fully connected layer followed by a Tanh activation, and a fully connected layer followed by a gated linear unit (GLU) layer \cite{dauphin2017language}. The $ n $-th MLP layer outputs the subband mask $\hat M_{n}$ with a shape of $ (2 \times C) \times T \times F_{n} $, which contains a mask of real values and a mask of imaginary values. All the outputs are concatenated along the frequency axis to derive a cIRM $ \hat{M} \in \mathbb{C}^{C \times T \times F} $.




\section{Experiments}\label{section:experiments}

This section first presents our implementation details of the winning system in Sound Demixing Challenge 2023 (SDX'23). Then, we describe the ablation study on the a smaller version of BS-RoFormer and the effectiveness of Rotary position embedding.

\subsection{Dataset}


The Musdb18HQ dataset \cite{rafii2017musdb18} contains 100 and 50 songs for training and evaluation, respectively. All recordings are stereo with a sampling rate of 44.1k Hz. Each recording contains four stems: vocals, bass, drums, and other. For MDX'23 submission, we also use an ``In-House'' dataset containing 500 songs: 450 songs are added to the training set, and 50 songs are for validation. Each song has four stems with a sampling rate of 44.1k Hz following \cite{rafii2017musdb18}. 


\subsection{Evaluation Metrics}
We adopt the signal-to-distortion ratio (SDR) \cite{vincent2006performance} implemented by \texttt{museval} \cite{SiSEC18} as the evaluation. The SDR score is calculated by:
\begin{equation} \label{eq:sdr}
SDR(y, \hat{y}) = 10 \log_{10} \frac{||y||^{2}}{||\hat{s} - s||^2}
\end{equation}
Higher SDR indicates better separation quality.

\subsection{Data Augmentation}

Our model takes a segment of 8-seconds waveform for input and output. For better training efficiency, we maintain a dynamic pool of 8-seconds stem-level segments in the memory. At each training step, the dataloader samples a batch of random 4-stem tuples from the pool. Each stem audio is processed with a random gain in a range of $\pm 3$ dB and has a chance of 10\% to be replaced with a silence waveform. The four stems of a tuple, which can originate from different songs, are mixed by linear addition to generate a training example. This random-mixing strategy will produce examples that are not musically aligned.

The dynamic pool has a size (e.g., 512 segments for each stem) larger than the batch size to facilitate wider diversity for a batch, and is updated in a first-in-first-out manner at each training step to remain its size.
To add samples to the pool, we crop an 8-seconds segment of 4 stems at a random moment from a full-length song. A stem-level segment must satisfy a loudness level larger than -50 dB. 



\subsection{Implementation Details}
\textbf{Band-split Module.}
We apply a Hann window size of 2048 and a hop size of 10 ms for STFT to compute the complex spectrogram of 8-seconds long input. We use the following band-split scheme: 2 bins per band for frequencies under 1000 Hz, 4 bins per band between 1000 Hz and 2000 Hz, 12 bins per band between 2000 Hz and 4000 Hz, 24 bins per band between 4000 Hz and 8000 Hz, 48 bins per band between 8000 Hz and 16000 Hz, and the remaining bins beyond 16000 Hz equally divided into two bands. This results in a total of 62 bands. All bands are non-overlapping. We derive this setting referring to \cite{luo2023music}. In our pilot study, slightly varying the band-slplit setting does not show significant difference in results. 

\vspace{0.1cm}
\noindent \textbf{Configuration.} 
We use $D$ = 384 for feature dimension, $L$ = 12 for the number of Transformer blocks, 8 heads for each Transformer, and a dropout rate of 0.1. The multi-band mask estimation module utilizes MLPs with a hidden layer dimension of $4D$. The resulting model has 93.4M parameters.
We use Pytorch-Lightning 2.0 and Exponential Moving Averaging (EMA) with a decay 0.999. For training, we adopt the AdamW optimizer \cite{loshchilov2017decoupled} with a learning rate (LR) $ 5 \times 10^{-4} $, and reduce LR by 0.9 every 40k steps. To optimize the GPU memory usage, we employ the checkpointing technique as well as the mixed precision, where the STFT and iSTFT modules use FP32 and all the others use FP16.


We trained three separation models respectively for vocals, bass, and drums using In-House and the Musdb18HQ training set. For the ``other'' stem, we subtracted the vocals, bass, and drums signals from the input mixture in the time domain. For each model, the training process lasted for 4 weeks using 16 Nvidia A100-80GB GPUs with a total batch size of 128 (i.e., 8 for each GPU). The model checkpoint with the best validation result was selected. 

\vspace{0.1cm}
\textbf{Enframe \& Deframe}. We use a hop size of 4 seconds for segment enframing. Two deframing methods are studied: ``truncate\&concat" (TC) and ``overlap\&average'' (OA). In TC, we discard the front and rear 2-seconds signals of each segment-level output and concatenate all truncated segments. In OA, each output segment has a 4-second overlap by the next segment at the rear, and the overlapped part is averaged.


\begin{table}[t]
\centering
\footnotesize The $\star$ symbol indicates the proposed BS-RoFormer.
\resizebox{\columnwidth}{!}{%
\begin{tabular}{lccccc}
 \toprule
 System  & Vocals & Bass & Drums & Other & Mean \\
 \midrule
 1. SAMI-ByteDance$^\star$ & 11.36 & 11.15 & 10.27 & 7.08 & 9.97 \\
 2. ZFTurbo \cite{solovyev2023benchmarks} & 10.51 & 9.94 & 9.53 & 7.05 & 9.26 \\
 3. kimberley\_jensen & 10.40 & 10.06 & 9.47 & 6.80 & 9.18 \\
 4. kuielab \cite{kim2023sound} & 10.01 & 9.72 & 9.43 & 6.72 & 8.97 \\
 5. alina\_porechina & 9.07 & 9.92 & 9.29 & 6.23 & 8.63 \\
 (Baseline) BSRNN \cite{luo2023music} & 7.98 & 5.63 & 6.53 & 4.43 & 6.14 \\
 \bottomrule
 \end{tabular}}
\caption{SDX'23 Leaderboard C final results (in global SDR).}
\label{table:mdx23}
\end{table}

\subsection{SDX'23 MSS Results}

SDX'23 MSS track featured three leaderboards, A: label noise, B: bleeding, and C: standard music separation. The first two leaderboards set some requirements about robustness, so the submitted systems can be trained only on the provided dataset. Our system participated the standard music separation track, where the submitted systems can be trained on any data with no limitation, so it is more desirable. Table \ref{table:mdx23} presents the final results on the organizer's private set for the top-5 systems of SDX'23 MSS leaderboard C \cite{fabbro2023sound}. We used TC deframing for the submission.

Our system (SAMI-ByteDance) outperforms the second best (ZFTurbo) by a large margin (0.71 dB on average), demonstrating the effectiveness of the proposed BF-RoFormer. When comparing the results of vocals, bass, and drums (since we did not train the ``other'' separator), the average SDR difference is 0.93 dB. In terms of listening experience, we found our model outputs are highly accurate, meaning they contain less residues from the background. However, the resulting spectrogram may look sharp and less foggy compared to that generated by existing CNN-based models, and this might not be a favorable feature to some audiences. According to \cite{fabbro2023sound}, our model outputs gained more preference from musicians and educators than from music producers in the listening test of SDX23. 




\begin{table}[t]
 \footnotesize
 The $\dag$ symbol indicates MSS systems trained with extra data. \hspace{21pt} \\ 
 The $\ddagger$ symbol indicates the results are evaluated on non-HQ version. \\
\resizebox{\columnwidth}{!}{%
\begin{tabular}{p{3.15cm}p{0.6cm}p{0.6cm}p{0.6cm}p{0.6cm}p{0.6cm}}
 \toprule
 & Vocals & Bass & Drums & Other & Avg. \\
 \midrule
 Conv-TasNet \cite{luo2019conv}$^\ddagger$ & 6.81 & 5.66 & 6.08 & 4.37 & 5.73 \\
 Spleeter \cite{hennequin2020spleeter}$^\dag$$^\ddagger$ & 6.86 & 5.51 & 6.71 & 4.55 & 5.91 \\
 ResUNet \cite{kong2021decoupling}$^\ddagger$ & 8.98 & 6.04 & 6.62 & 5.29 & 6.73 \\
 HDemucs \cite{defossez2021hybrid}$^\dag$ & 8.13 & 8.76 & 8.24 & 5.59 & 7.68 \\
 BSRNN \cite{luo2023music} & 10.01 & 7.22 & 9.01 & 6.70 & 8.24 \\
 BSRNN 
 \cite{luo2023music}$^\dag$ & 10.47 & 8.16 & 10.15 & 7.08 & 8.97 \\
 Sparse HT Demucs \cite{rouard2023hybrid}$^\dag$  & 9.37 & 10.47 & 10.83 & 6.41 & 9.27 \\
 \midrule
 BS-Transformer ($L$=6, OA)  & 9.15 & 6.11 & 3.08 & 4.77 & 5.78 \\
 BS-RoFormer ($L$=6, TC)  & 10.68 & 11.28 & 9.41 & 7.68 & 9.76 \\
 BS-RoFormer ($L$=6, OA)  & 10.66 & 11.31 & 9.49 & 7.73 & 9.80 \\
 BS-RoFormer ($L$=12, OA)$^\dag$  & 12.72 & 13.32 & 12.91 & 9.01 & 11.99 \\
 \bottomrule
 \end{tabular}}
\centering
\caption{Median SDRs of different MSS models.}
\label{table:is_baseline}
\end{table}

\subsection{Ablation Study}

We investigate the effects of three aspects: 1) the number of Transformer blocks; 2) RoPE or absolute positional encoding; 3) TC or OA deframing methods. To this end, we implement three smaller variants of the proposed model with $L$=6. In one variant, we remove the RoPE and add learnable absolute positional embeddings in the attention module, and call it ``BS-Transformer,'' since it uses a standard Transformer. We train a dedicated separation model for the ``other'' stem, except ``BS-RoFormer ($L$=12, OA),'' which is our MDX'23 submission. The numbers of parameters for BS-RoFormer and BS-Transformer with $L$=6 are 72.2M and 72.5M, respectively. Models with $L$=6 are trained solely on the Musdb18HQ training set using 16 Nvidia V100-32GB GPUs. We do not use the In-House dataset for ablation study. The effective batch size is 64 (i.e., 4 for each GPU) using accumulate\_grad\_batches=2.

Table \ref{table:is_baseline} presents the comparison between our proposed models and existing models. We report the median SDR across the median SDRs over all 1 second chunks of each test song in Musdb18HQ following prior works. First, BS-RoFormer with $L$=6 is still very competitive and can achieve state-of-the-art performance compared to models trained without extra training data. ``BS-RoFormer$^\dag$ ($L$=12, OA)'' outperforms all existing models by a large margin (over 2 dB on average). Second, BS-Transformer without RoPE does not seem to work given the low SDRs, demonstrating that RoPE is crucial in our proposed architecture as discussed in Section \ref{sec:attention}. According to our observations, the training progress of BS-Transformer is very slow, and it still remains low SDRs after two weeks of training on Musdb18HQ. Instead, BS-RoFormer models with $L$=6 get converged within a week. Lastly, the OA deframing shows better performance than TC except vocals. Qualitatively, OA offers smoother song-level quality that can improve the overall listening experience.

\section{Conclusion}\label{section:conclusion}
We have presented the BS-RoFormer model, which is based on a novel hierarchical RoPE Transformer architecture. Its outstanding performance may shed the light on the development of next generation MSS systems.
For future work, we would focus on improving the qualitative performance. 
The sharp sound quality may be improved by introducing overlapping band projection at the front-end.

\small
\bibliographystyle{IEEEbib}
\bibliography{refs}

\end{document}